\documentclass[twocolumn,showpacs,citeautoscript,floatfix]{revtex4}
\usepackage{amssymb}
\usepackage{graphicx}
\usepackage{dcolumn}   
\usepackage{bm}        
\usepackage[]{amsmath}

\begin{document}

\title{ Spin-interactions in Mineral Libethenite Series: Evolution of Low-dimensional Magnetism}

\author{Debjani Karmakar}

\affiliation{ Technical Physics Division, Bhabha Atomic research Center, Mumbai 400085, India.}

\author{J. V. Yakhmi}

\affiliation{ Technical Physics Division, Bhabha Atomic research Center, Mumbai 400085, India.}
\date{\today}
\begin{abstract}
Interesting magnetic properties and spin-exchange interactions along various possible pathways in half-integral spin quantum magnetic  tetramer system: A$_{2}$PO$_{4}$OH (A = Co, Cu) are investigated. Interplay of structural distortion and the magnetic properties with the evolution of localized band structure explain the gradual transition from a three-dimensional antiferromagnet to a low-dimensional frustrated magnetic system along the series. Detailed study of exchange mechanism in this system explores various possibilities of complex magnetic interaction. Electronic structure of this series, studied with the help of different appropriate density-functional approaches like N-th order Muffin-Tin orbital(NMTO) and Planewave Pseudopotential calculations incorporating onsite Coulomb repulsion(U), identifies the underlying magnetic exchange mechanism of this series. Thereafter a generalized minimal model spin-Hamiltonian is constructed for the low-dimensional system. Solution of this model Hamiltonian within first order perturbation theory results into the evaluation of spin-gap in the spin-tetramer system. In addition, the effects of size-confinement and volume reduction on the relevant exchange integrals and spin-gap of the low-dimensional system are also discussed.
\end{abstract}
\maketitle
\section{Introduction}
Thorough understanding of the interrelated structural and magnetic properties of complex geometrically frustrated systems remain always interesting as absence of long-range magnetic order in such systems opens up the possibilities of a wide range of excited state properties \cite{vasilev}. Such half-integral spin, especially the spin-1/2 low-dimensional magnetic system with a singlet ground state, attracts much attention due to their fundamental quantum nature \cite{ueda, bose} and are also considered to be suitable two-level system for quantum computing.\cite{mayaffre} Observation of pseudo spin-gap in high-$T_{c}$ cuprates \cite{itoh} has motivated intensive study of spin-exchange mechanism of new spin-gap systems. Increasing experimental activity on complex low-dimensional magnetic systems \cite{shin1,shin3, cepas, matsuda} motivates a detailed theoretical first-principles investigations of the underlying mechanism of substitution-induced structural distortions and the resulting changes in magnetic exchange properties to understand the evolution of a long-range ordered system into a low-dimensional one. Moreover, for complex geometrically frustrated system, identification of exchange paths relevant to the formation of minimal spin-Hamiltonian and thus determination of effective magnetic mechanism can provide the insight about magnetostructural correlations.

A$_{2}$PO$_{4}$OH (A = Co, Cu), the natural mineral Libethenite, possessing emerald green and dark purple colour for Cu and Co based systems respectively, belongs to the phosphate mineral class \cite{harrison} and is mostly used for their widespread catalytic properties. \cite{pedro1}  The parent compound, Co$_{2}$PO$_{4}$OH is experimentally observed to possess a three-dimensional antiferromagnetic order at around 70 K. Substitution of Cu gradually in place of  Co1, Co2 and finally for both Co1 and Co2 sites leads to the onset of low-dimensional magnetic properties in this system. \cite{pedro1, pedro} The final compound Cu$_{2}$PO$_{4}$OH, a spin-1/2($d^9$) low-dimensional tetrameric system, being the most interesting among such class, will be investigated in more detail in the present work. Previous experimental observations on Cu$_{2}$PO$_{4}$OH system \cite{belik, kuo} predict a singlet ground state with a spin-gap $\sim $141 K. A simpler analogue of the spin-exchange mechanism were put forward via the square spin-tetramer(SQST) model.\cite{kuo} In the present study, we adopt suitable Density Functional Theory (DFT)-based  methodologies to first identify and then calculate the relevant exchange interactions in such systems, which appears to be more complicated than the SQST model.\cite{kuo} Next, we construct the minimal Heisenberg Model for the final system and approximately solve it within first order perturbation theory to evaluate the spin-gap between the singlet ground state and the triplet excited state. Such multistep analysis will be helpful to build a generalised spin-model Hamiltonian for the low-dimensional tetramer systems and also provides understanding of the correlation between the magnetic and structural properties. For the last system, effects of size-reduction and cell-volume reduction on the exchange mechanism are also studied.
\section{First principles analysis}
A$_{2}$PO$_{4}$OH belongs to the space group Pnnm(58) with two types of the transition metal cation A (A1 or A2, suitably replaced by Cu or Co) and four types of O-ligand(O1,O2,O3 and O4). A1 forms a distorted octahedra with O1, O2 and O4(H), whereas A2 is fivefold coordinated with O1, O3 and O4(H)(Fig 1 and 5(a)). The experimental value of lattice parameters are obtained from reference \cite{harrison}. For a detailed understanding of magnetic interactions in the tetrameric system, we have investigated all four systems in the series, \textit{viz}. (I) Co$_{2}$PO$_{4}$OH, (II) CuCoPO$_{4}$OH(Cu replacing Co1), (III) CoCuPO$_{4}$OH (Cu replacing Co2) and (IV) Cu$_{2}$PO$_{4}$OH, all generated from the parent Libethenite system I. First principles Density functional investigation of this series involves three steps; \textit{viz}. (1) For all four systems, lattice parameters and the atomic positions in the 36-atom unit cell are optimized using conjugate-gradient relaxation criteria as implemented in VASP \cite{vasp1,vasp2} with projector augmented wave(PAW)\cite{blochl} formalism. Bond-length and bond-angles of the relaxed structure provide a prior idea of the effects of structural distortion on the electronic and magnetic properties due to Cu substitution;(2) With the optimized positional coordinates and lattice parameters as input, we have utilized N-th order Muffin-Tin orbital (NMTO) downfolding technique to obtain the tight-binding hopping parameters for the Cu and Co localized \textit{d}-bands. Under second-order perturbation theory, computation of these hopping paramaters (\textit{t}$_{i}$)corresponding to some specific paths may lead to a quantitative comparison of the exchange couplings along them. For antiferromagnetic couplings, $4\textit{t}_{i}^2 / \textit{U}$ is a measure of the exchange coupling \textit{J}$_{i}$, with \textit{U} being the onsite Coulomb interaction strength for Co and Cu - 3\textit{d} bands; (3) After obtaining these relevant exchange paths, we have estimated the exchange coupling constants and the nature of exchange along these pathways by a total energy calculation for different spin-configurations in various supercell configurations consisting of 72 atoms within Local Spin-density Approximation(LSDA) + onsite Coulomb repulsion(U)formalism with VASP-PAW potentials.
\begin{figure}
{\includegraphics[width=45mm,clip]{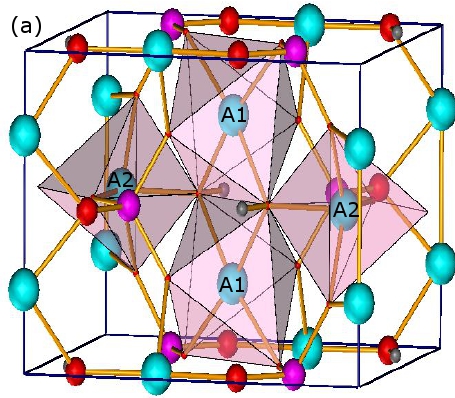}}
{\includegraphics[width=40mm,clip]{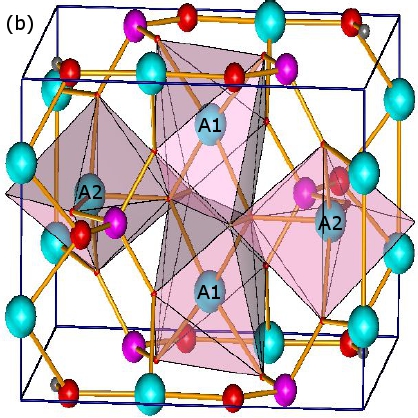}}
\caption{\label{fig1a} (colour online)The colour code of the atoms are Cu(Co) - Cyan, O - red, P - magenta, H - grey.(a): Corner shared A1O$_{6}$ octahedra and A2O$_{5}$ trigonal bipyramids are shown for comparison with the actual structure.(b): Actual atomic coordination showing A1O$_{4}$(OH)$_{2}$ edge-sharing infinite octahedral chains along the \textit{z}-axis and A2O$_{4}$(OH) edge-sharing trigonal bi-pyramidal dimers.}
\end{figure}
\begin{figure*}
{\includegraphics[width=170mm,clip]{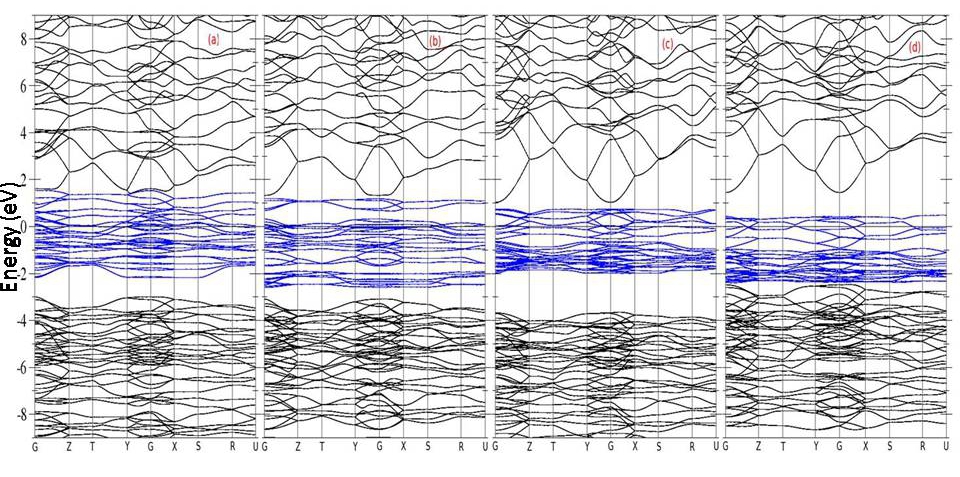}}
\caption{\label{fig2a}(color online) The figure represents the LSDA band structure for all four systems (a)Co$_{2}$PO$_{4}$OH, (b)CuCoPO$_{4}$OH(Cu replacing Co1), (c) CoCuPO$_{4}$OH (Cu replacing Co2) and (d) Cu$_{2}$PO$_{4}$OH with the localized NMTO-\textit{d}-bands near Fermi-level in blue. Band-width reduction of the localized manifold from system I and II to III and IV can be observed.}
\end{figure*}
The first step consists of the analysis of optimized structure. With gradual substitution of Cu in the parent Co$_{2}$PO$_{4}$OH system, the relaxed structure indicates that due to smaller size of Cu$^{2+}$ ion, there is sizable amount of structural distortions including change in the bond angle and bond-length around the site of substitution. The trend of significant bond-length, ionic distance and bond-angle modifications are presented in table I for comparison.
From system I to IV, as a combined effect of decrease in O2-A1-O2 and O1-A1-O1 and increase in A1-O2-A1 bond-angles, the difference between the intra and inter tetramer A1-A1 \textit{z}-direction distance nullifies along the series (Table I) and thereby the \textit{z}-direction A1-A1 dimer chain structurally turn into a monomer chain. This also leads to a Jahn-Teller like elongation of the distorted octahedra. Although the A1-O1-A2 angle does not change much, but as a result of increase in A1-O1-A1 and A2-O3-A2 angle, the intra-tetramer A1-A2 distance reduces and inter-tetramer A2-A2 distance increases along the series. Interestingly, disparity of bond-lengths of A1 and A2 with O1 and O2 and H-ligands leads to a competition in the cationic covalency which also has footprints on the ligand-field splitting and localized band structure as discussed in the next step. However, system (III) slightly breaks the gradual order in some structural details, viz, A1-O1-A1 and A2-O3-A2 bond angles are more than system IV, which leads to a change in the coordination of ligand O4 around A1 (dist $\sim$ 1.98{\AA}) such that A1 - O4 bonds are formed for system III.
\begin{table}[tb*]
\caption{\label{tab1} Important intra and inter-tetramer distances (d) and transition-metal cation to ligand bond-lengths (b)and bond-angles for the relaxed structures of the Libethenite-series. See fig 5(a) for structure.}
\begin{ruledtabular}
\begin{tabular}{ccccc}
d({\AA})&I & II & III & IV   \\
\hline
A1-A2(intra) & 3.47 & 3.43 & 3.35 & 3.30  \\
A1-A1(intra) & 2.78 & 2.82 & 2.90 & 2.95  \\
A1-A1(inter) & 2.89 & 2.91 & 2.89 & 2.95  \\
\hline
b({\AA})& I & II &  III & IV   \\
\hline
A1-O1 & 1.91 & 1.93 & 2.00 & 1.97   \\
A2-O1 & 2.03 & 1.97 & 1.87 & 1.93  \\
A2-O3 & 1.95 & 1.93 & 1.99 & 2.14  \\
A1-O4 & 2.28 & 2.40 & 1.98 &  2.32  \\
\hline
Angle(degrees) & I & II & III & IV  \\
\hline
O2-A1-O2 & 96.52 &	90.35 &	 86.88 & 82.47 \\
O1-A1-O1 & 83.88  & 83.03 &  82.50 & 81.20  \\
A1-O2-A1 & 95.99 & 96.64 & 97.99 &  98.40 \\
A1-O1-A1 & 96.11 & 96.33 & 98.09 &  97.35 \\
A2-O3-A2 & 99.71 &  100.78& 102.82 & 101.42  \\
A1-O1-A2 &	123.40 & 122.86 & 122.30  & 121.64 \\
\hline
\end{tabular}
\end{ruledtabular}
\end{table}

Incorporation of H-ion into the ligand coordination significantly changes the ligand field and the resulting splitting for this series of materials. The three-dimensional real structure of the system consists of A1O$_{4}$(OH)$_{2}$ \textit{plane-sharing} infinite octahedral chains along the \textit{z}-axis, \textit{edge-shared} with A2O$_{4}$(OH) trigonal bi-pyramidal dimers. This structure is distorted from the \textit{edge-shared} A1O$_{6}$-octahedra, \textit{corner-shared} with A2O$_{5}$ trigonal bipyramids as shown in fig 1(a). Figures 1(a) and (b) depict a comparison of the distorted ligand coordination around A1 and A2 due to H-ion. The ligand field splitting for A1 and A2 3\textit{d}-levels are also deviated from the usual octahedral and trigonal-bipyramidal ones and becomes more closer to a distorted square-antiprismatic (coordination no. 8) and octahedral (coordination no. 6)coordination for A1 and A2, respectively. A closer look to fig 1(b) also reveal that actual coordination around A2 is distorted octahedral and two plane shared octahedra around A1 along \textit{z} resembles more of a distorted antiprism.  This description will be more evident in the orbital-analysis of the different 3\textit{d}-characters.

In the second step, the relaxed structural inputs are used to extract the resulting effects of structural distortion on the localized bandstructure. Orbital analysis of the localized bands provides an idea about the connection between the change of ligand coordination and ligand field splitting with the magnetic properties along the series. Fig 2(a)-(d) represent the LSDA bands of the relaxed structures with the A1 and A2-3\textit{d} localized N-th order muffin-tin orbital (NMTO)-downfolded bands \cite{nmto1, nmto2} (marked in blue) around Fermi-level.  Localized 3\textit{d}-bandwidths for these four systems are 3.71 eV, 3.76 eV, 2.72 eV and 2.8 eV, respectively. This reduction in bandwidth of $\sim1 eV$ from the first two to the last two systems can be mostly attributed to the structural distortion. The orbital projected Co and Cu 3d-fatbands for systems I and IV are presented in fig. 3 and 4 respectively. The nature of orbital character distribution for system II and III are similar to I and IV. The labelling of various atoms can be seen from fig 5(a).

Usually, for low-dimensional Cu$^{2+}$-based dimer systems, \cite{azurite, sarita} eight Cu-3\textit{d}$_{x^{2} - y^{2}}$ bands localize near E$_{F}$. In the present case, highly localized A1 and A2 3$d_{x^{2} - y^{2}}$ bands for system I and II also implies the presence of dimer-interactions. For system III and IV, however, the dimer like localization of A1 and A2- 3\textit{d}$_{x^{2} - y^{2}}$ levels smears out due to hybridization with other orbitals indicating existence of more complex magnetic interaction. For A1, the bonding orbitals are $x^{2} - y^{2}$ and 3$z^{2} -1$, while \textit{xy}, \textit{yz} and \textit{xz} have both bonding and antibonding contributions. The situation is just reversed for A2. A1- 3$d_{3z^{2} -1}$ orbitals form sigma-bond with O1 and O2 \textit{p}$_{z}$, rendering the respective bands to be the lowest in energy for all systems. A1-3\textit{d}$_{x^{2} - y^{2}}$ is \textit{pi}-bonded with O1 and O2 \textit{p}$_{x}$ and \textit{p}$_{y}$ and thus gets localized at a lower energy level than \textit{xy}. For A1, \textit{xz} and \textit{yz} orbitals are of highly hybridized in nature, the bonding part of which originates from the overlap of \textit{z}-branch of \textit{xz} or \textit{yz} with O1 and O2-\textit{p}$_{z}$. The antibonding part is from the \textit{x} or \textit{y} branch, which due to H-O1 charge transfer, have little opportunity to interact with O1-\textit{p}$_{x}$ or \textit{p}$_{y}$. The occupied part of A1-3\textit{d}$_{xy}$ originates due to the overlap with O4-\textit{p}$_{x}$ and \textit{p}$_{y}$. For A2, 3\textit{d}$_{xy}$ and \textit{x} and \textit{y} branch of 3\textit{d}$_{xz}$ and 3\textit{d}$_{yz}$ is sigma-bonded with O3-\textit{p}$_{x}$ and \textit{p}$_{y}$. 3\textit{d}$_{x^{2} - y^{2}}$ and 3$d_{3z^{2} -1}$ is mostly of antibonding nature. However, for the third system, additional covalency with O4-\textit{p} lowers both of the orbital energies, especially for the 3$d_{3z^{2} -1}$.
\begin{figure*}
{\includegraphics[width=110mm,clip]{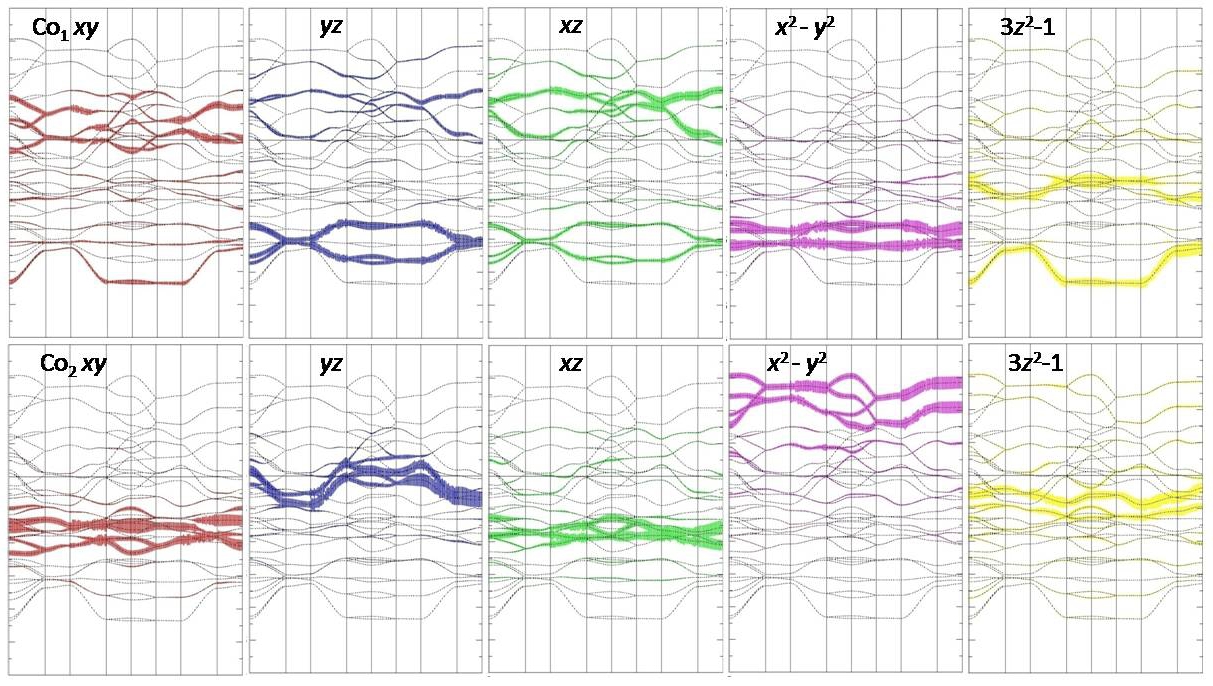}}
\caption{\label{fig2a} (colour online)The figure represents the orbitally projected NMTO fatbands for Co1-d and Co2-d bands for the system Co$_{2}$P$O_{4}$OH. The high-symmetry points are G, Z, T, Y, G, X, S, R, U, similar to the bands presented in fig 2. The energy-axis ranges from -3 to +2 eV at an interval of 0.5 eV.}
\end{figure*}
\begin{figure*}
{\includegraphics[width=110mm,clip]{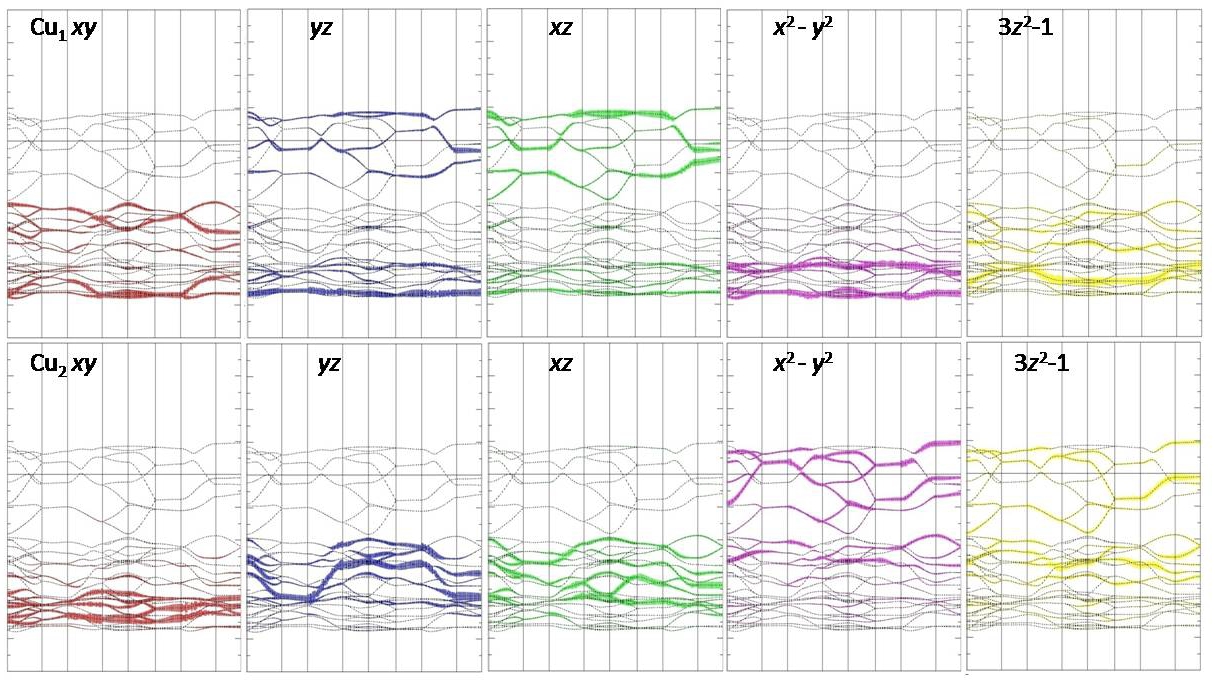}}
\caption{\label{fig3a} (colour online)The figure represents the orbitally projected NMTO fatbands for Cu1-d and Cu2-d bands for the system Cu$_{2}$P$O_{4}$OH. The high-symmetry points are G, Z, T, Y, G, X, S, R, U, similar to the bands presented in fig 2. The energy-axis ranges from -3 to +2 eV at an interval of 0.5 eV.}
\end{figure*}
In the present series, competitive cationic covalency effects are prominent from the orbital analysis. For the same O1 \textit{p}$_{x}$ or \textit{p}$_{y}$ orbitals, there is a competition between A1- 3$d_{x^{2} - y^{2}}$, A2-3\textit{d}$_{xy}$ and H-1\textit{s}.  For system I and II, A1-O1 bond-length is less than A2-O1 bond (Table I) indicating greater chances for A1- 3$d_{x^{2} - y^{2}}$ covalency. Thus for A1,  3$d_{x^{2} - y^{2}}$ bands localize at a lower energy compared to A2. For systems III and IV, A1-O1 bond-length is greater than A2-O1 bond. Thus, A2- 3$d_{x^{2} - y^{2}}$ levels stay at lower energy compared to the first two systems. For the third system, additional lowering occurs due to bonding with O4-\textit{p}$_{x}$ and \textit{p}$_{y}$. Also, for system III, strong interaction of \textit{d}-bands with \textit{p}$_{x}$ and \textit{p}$_{y}$ orbital of O3 and O4 and \textit{p}$_{z}$ orbital of O1 and O2 pushed the filled ligand levels far below (around -4 eV) in comparison to the other systems(Fig 2).
Investigation of hopping parameters along relevant paths with the help of NMTO calculations identifies six significant couplings for such system, \textit{viz}. the intra-tetrameric exchange $J_{1}$ and $J_{1}^{\prime}$ between A1 and A2, the dimeric exchange $J_{2}$ and $J_{2}^{\prime}$ along \textit{z}-axis A$_{1}$ - array and another in-plane dimeric exchange $J_{3}$ and $J_{4}$ between A$_{2}$. Figure 5(a) shows the exchange paths with arrows along with the structure and figure 5(b) depicts the schematic diagram of predicted exchange couplings for this system.
\begin{figure}
{\includegraphics[width=70mm,clip]{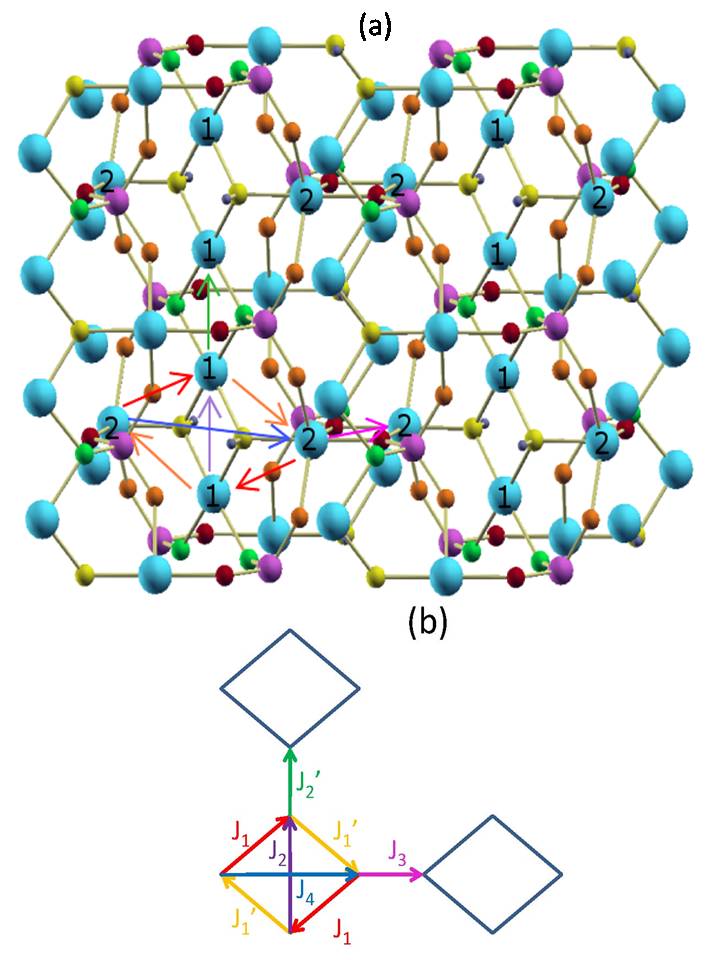}}
\caption{\label{fig1a}(colour online)(a):The three-dimensional structure of one of the libethenite compound (IV) Cu$_{2}$PO$_{4}$OH, with the relevant exchange paths denoted with arrows of different colours. The red and orange coloured arrow represents $J_{1}$ and $J_{1}^{\prime}$(see text). The \textit{z}-direction dimeric exchanges are $J_{2}$(violet) and $J_{2}^{\prime}$(green) with in-plane dimeric exchanges between Cu2 $J_{3}$ (magenta) and $J_{4}$ (blue). Different O-ligands are coloured as O1(yellow), O2(green), O3(red) and O4(orange) (b): Schematic diagram of important exchange paths.}
\end{figure}

After identifying these relevant exchange paths, in the next step, we have estimated the exchange coupling constants and the nature of exchange along these pathways by a total energy calculation for different spin-configurations in various supercell configurations consisting of 72 atoms. Under LSDA + U approximation, the total energy for each spin configuration is calculated with VASP-PAW potentials after ionic relaxation. The significant exchange coupling constants for all four systems are listed for the optimum \textit{U} (5 eV for Cu and 4.5 eV for Co) and \textit{J }(1 eV for Cu and 0.9 eV for Co) value in Table II. The insulating nature of all these four compounds are correctly represented with LSDA + U calculation.
For the system I (table II), the tetrameric exchange is actually composed of two kinds of dimeric exchanges $J_{1}$ and $J_{1}^{\prime}$ along the two Co1 - O1 - Co2 paths (Fig 5(a)). Along the \textit{z}-direction, the Co1 - O2 - Co1 ($J_{2}$) and Co1 - O1 - Co1 ($J_{2}^{\prime}$) also forms a dimer. The in-plane-Co2-dimer comprises of Co2 - O3 - Co2 ($J_{3}$) and Co2 - O1 - O1 - Co2 ($J_{4}$) exchanges. The values of $J_{1}$, $J_{1}^{\prime}$, $J_{2}$, $J_{2}^{\prime}$ (table II) are of the same order and AFM. The Co2-dimeric exchange $J_{3}$ is FM and $J_{4}$ is AFM. Since all the couplings are not AFM, instead of frustration, long-range magnetic order prevails in this system. The order of magnitude of the couplings also implies that the magnetic order is three dimensional which appears to be at par with the experimental results.\cite{pedro} With gradual substitution of Cu in place of Co1 and Co2 in system II and III, $J_{1}$ and $J_{1}^{\prime}$ becomes more and more isotropic, the dimeric system will gradually turn into a tetrameric one. Also the \textit{z}-direction dimer turn into a monomer chain($J_{2}= J_{2}^{\prime}$). Although most of the exchanges for all the three other systems are AFM, for system III, exchanges along some paths are strongly ferromagnetic, which may lead to a long range ferromagnetic behaviour. For the final system, Cu$_{2}$PO$_{4}$OH, Cu1 - O1 - Cu2 exchange is exactly tetrameric ($J_{1}$ = $J_{1}^{\prime}$) with \textit{z}-direction Cu1 - Cu1 monomer($J_{2}= J_{2}^{\prime}$)and in-plane Cu2 - Cu2 dimers. All nearest neighbour couplings are AFM (Table II)as a result of geometric frustration in this particular system. The monomer and dimer exchange values are less than the tetrameric exchange implying the onset of low-dimensional magnetic behaviour in this system. The magnetostructural details of this series are in resemblance with the experimental studies performed in reference \cite{pedro1}.
\begin{table}[tb*]
\caption{\label{tab2} Relevant exchange couplings constants in meV for the series of mineral libethenites.}
\begin{ruledtabular}
\begin{tabular}{ccccccc}
System & $J_{1}$ & $J_{1}^{\prime}$ & $J_{2}$ & $J_{2}^{\prime}$ & $J_{3}$ & $J_{4}$  \\
Co$_{2}$PO$_{4}$OH & -25.84 & -25.2 & -22.8 & -23.07 & 25.34 & -10.56 \\
CuCoPO$_{4}$OH & -22.23 & -24.47 & -15.92 & -13.6 & 12.43 & -10.7 \\
CoCuPO$_{4}$OH & -18.02 & -19.04 & 22.07 & 22.14 & 20.77 & -3.62 \\
Cu$_{2}$PO$_{4}$OH & -20.30 & -20.30 & -10.12 & -10.12 & -9.15 & -2.18 \\
\end{tabular}
\end{ruledtabular}
\end{table}

A large number of analytical investigations are available for the spin-chain low-dimensional systems \cite{mg1,mg2,haldane1,haldane2,gayen,partha}and also for more complicated systems like spin-trimers, tetramers,\cite{dadi, trivedi} octahedra \cite{indrani} \textit{etc}. We may construct a generalized minimal Heisenberg Model spin-Hamiltonian describing the low-energy spin-interactions and excitations  for the low-dimensional tetramer system as:
\begin{eqnarray}
H=J_{1}\sum_{i,j=1, i\neq j}^{4}\textbf{S}_{i}.\textbf{S}_{j}+J_{2}\sum_{i=1}^{n}\textbf{S}_{2i-1}.\textbf{S}_{2i+1}\notag \\
+(J_{3}+J_{4})\sum_{i=1}^{n}\textbf{S}_{2i}.\textbf{S}_{2i+2}
\end{eqnarray}
Here \textit{n} is the number of tetramer units. For the isolated spin-cluster as presented in Fig 5(b), equation (1) can be simplified as:
\textit{H} = \textit{J}$_{1}$(\textit{\textbf{S}}$_{1}$.\textit{\textbf{S}}$_{2}$ + \textit{\textbf{S}}$_{2}$.\textit{\textbf{S}}$_{3}$ + \textit{\textbf{S}}$_{3}$.\textit{\textbf{S}}$_{4}$ + \textit{\textbf{S}}$_{4}$.\textit{\textbf{S}}$_{1}$) + 3\textit{J}$_{2}$ (\textit{\textbf{S}}$_{3}$.\textit{\textbf{S}}$_{1}$) + 2\textit{J}$_{3}$(\textit{\textbf{S}}$_{2}$.\textit{\textbf{S}}$_{4}$) + \textit{J}$_{4}$(\textit{\textbf{S}}$_{2}$.\textit{\textbf{S}}$_{4}$).
Within first order approximation, the degenerate $|S_{ztot}|=1$ triplet states for this system are given by $1/\sqrt{2}(|\uparrow\uparrow\uparrow\downarrow\rangle -|\uparrow\downarrow\uparrow\uparrow\rangle)$, $1/\sqrt{2}(|\downarrow\downarrow\downarrow\uparrow\rangle-|\downarrow\uparrow\downarrow\downarrow\rangle)$, $1/\sqrt{2}(|\uparrow\uparrow\downarrow\uparrow\rangle-|\downarrow\uparrow\uparrow\uparrow\rangle)$ and $1/\sqrt{2}(|\downarrow\downarrow\uparrow\downarrow\rangle-|\uparrow\downarrow\downarrow\downarrow\rangle)$. The $|S_{ztot}|=0$ degenerate singlet states are given by $1/\sqrt{2}(|\uparrow\uparrow\downarrow\downarrow\rangle-|\downarrow\downarrow\uparrow\uparrow\rangle)$ and $1/\sqrt{2}(|\uparrow\downarrow\downarrow\uparrow\rangle-|\downarrow\uparrow\uparrow\downarrow\rangle)$. Calculation of eigenvalues corresponding to these eigenstates leads to evaluation of the spin-gap, \textit{i.e}., the energy difference between this singlet ground state and triplet excited states. As a next step, we will investigate the effect of volume reduction on the magnitude of the spin-gap using this simple model Hamiltonian.
\begin{figure}
{\includegraphics[width=90mm,clip]{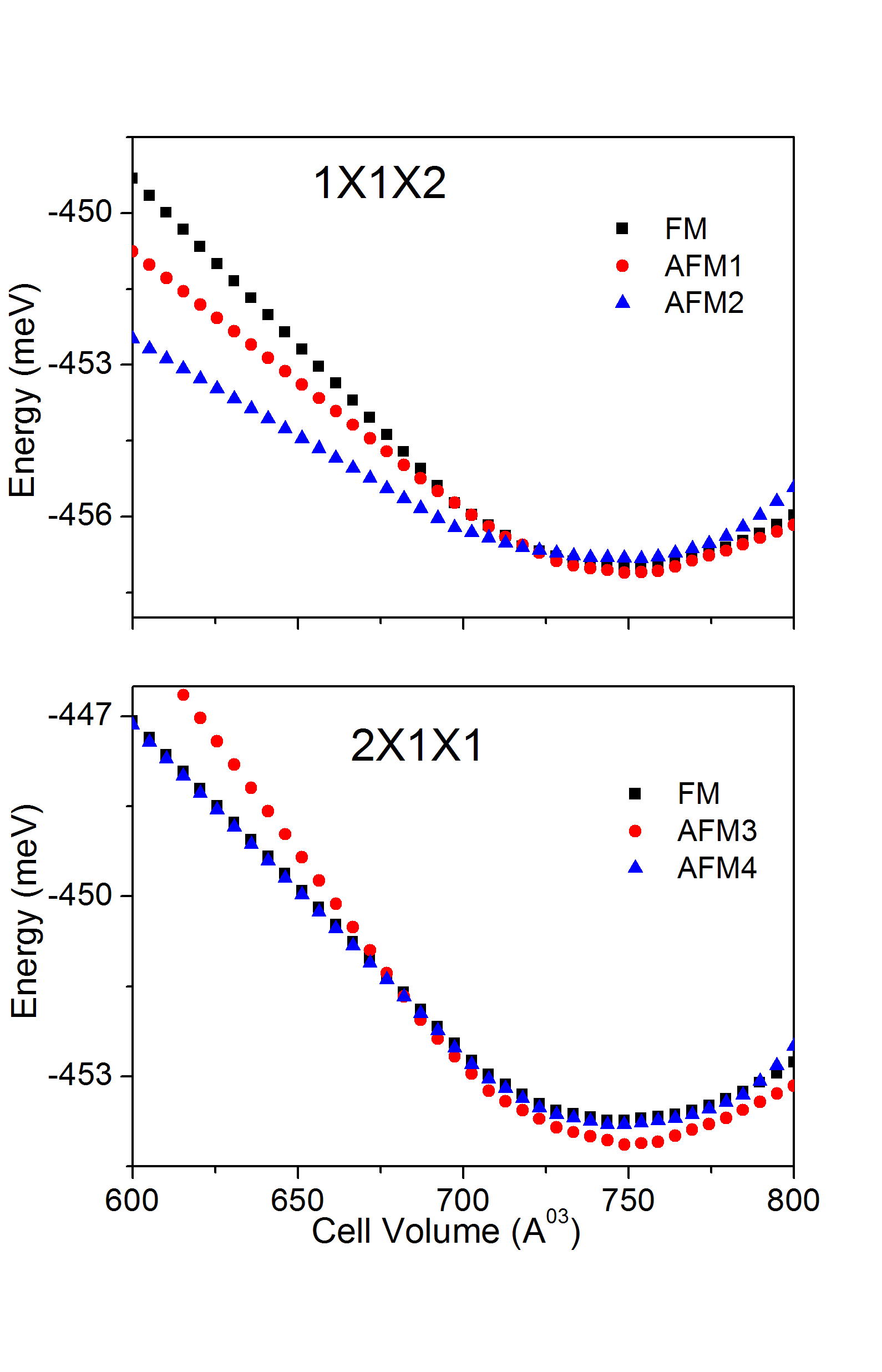}}
\caption{\label{fig3a} (colour online)Effect of cell volume reduction is equivalent to application of hydrostatic pressure. upper and lower panel represent the comparison of lowest energy configuration for $1\times1\times2$ and $2\times1\times1$ supercells (details in text).}
\end{figure}

\begin{figure}
{\includegraphics[width=90mm,clip]{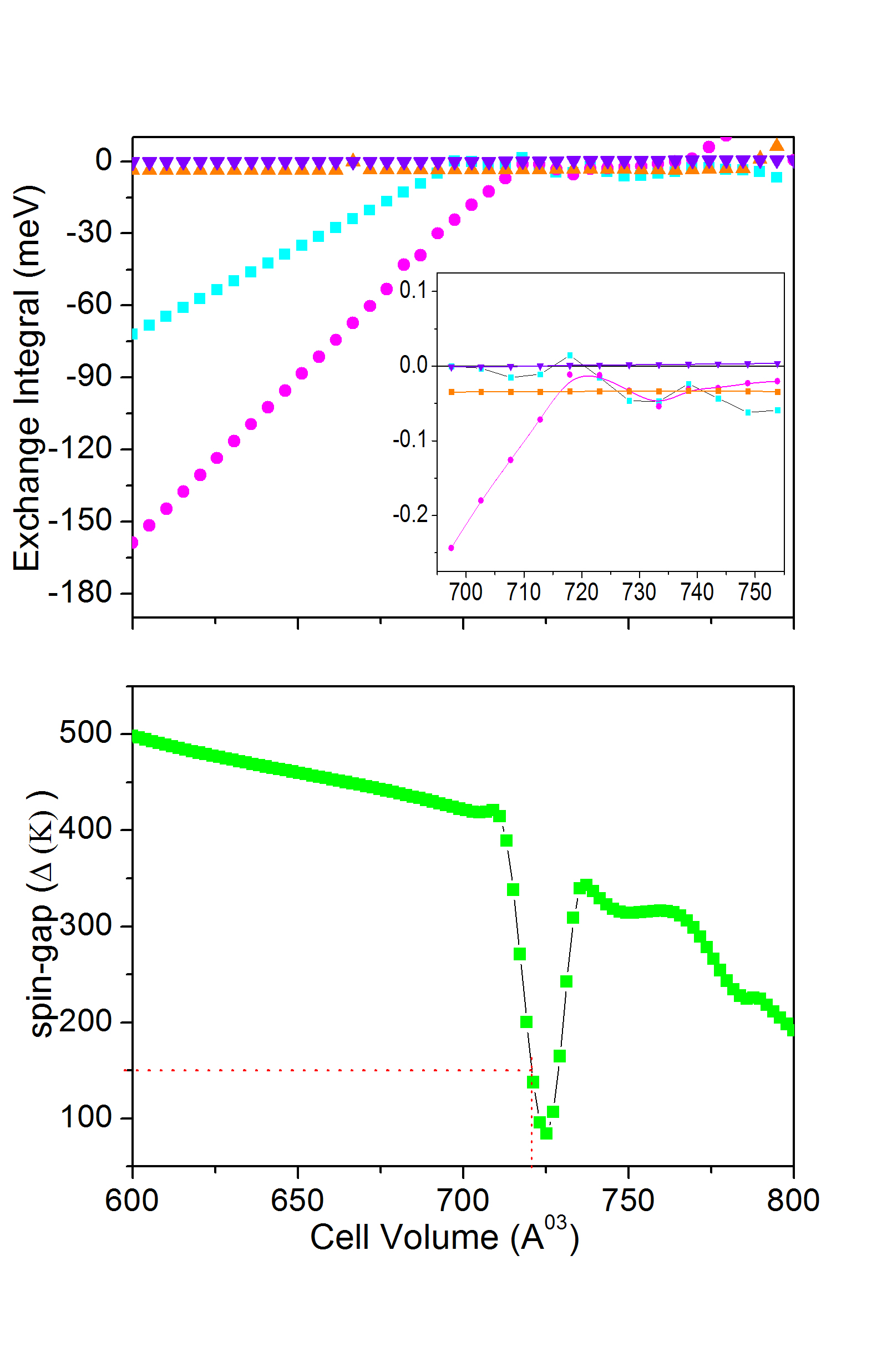}}
\caption{\label{fig3a} (colour online) Upper panel depicts the variation of exchange couplings with cell-volume in a system where all four couplings are present. The lower panel represents the variation of spin-gap values (in K) with cell-volume as computed numerically from the simplified model Hamiltonian (2). The red-line in the figure shows the equilibrium cell-volume and the corresponding spin-gap value $\sim$ 150 K.}
\end{figure}
\section{Investigation on spin-gap}
Among these four systems, presence of spin-gap is experimentally confirmed only for Cu$_{2}$PO$_{4}$OH. Hence, we have studied this particular system under volume reduction (equivalent to the application of hydrostatic pressure) and size confinement. To study the volume reduction effects, both the lattice parameters and ionic coordinates of various supercell configurations are relaxed with a particular magnetic arrangement.  For a $1\times1\times2$ supercell, FM denotes the configuration where all the Cu-atoms are ferromagnetically alligned. In AFM1, only the intra-tetrameric exchange $J_{1}$ is present. In AFM2, the monomeric exchange $J_{2}$ is also present. With gradual volume-reduction, AFM2 becomes the lowest energy configuration for $1\times1\times2$ supercell, as seen at the upper panel of fig 6. Similarly, for a $2\times1\times1$ supercell, AFM3 is the magnetic configuration where Cu2 - Cu2 dimeric exchange $J_{3}$ is present alongwith $J_{1}$. In AFM4, all of $J_{1}$,  $J_{3}$ and $J_{4}$ are present. In this case, AFM4 becomes the lowest energy configuration(fig 6, lower panel)with volume reduction with a very small energy difference with FM configuration. Thus, with gradual cell-volume reduction, the monomeric and dimeric exchanges become more and more important in this system. In addition, we have also observed that relaxation of the system in different magnetic configuration results into slightly different equilibrium volumes of the supercell.

Evolution of these four effective exchange integrals with reduced volume, as presented at the upper panel of Figure 7, are computed for a $2\times2\times2$ supercell, where all sorts of magnetic configurations are possible. The exchange couplings are plotted with respect to an average cell volume of 72 atoms, to ease a comparison with Figure 6. The equilibrium cell volume is also computed in a similar way. This figure indicates that with reduced volume, all of the effective couplings become more and more antiferromagnetic, with J$_{3}$ and J$_{4}$ orders of magnitude smaller than the other two. Variation of the spin-gap(in K) with volume, as calculated with the simplified model spin-Hamiltonian (in text after equation 1), is also plotted at the lower panel of Fig 7. It is evident from the figure that with all four effective couplings present, a decrease from the equilibrium cell volume (denoted by red dotted line in the figure) may result into an increase in the spin-gap magnitude. The situation is slightly different with an increase of cell volume, where the spin-gap first decreases and then increases. The marked red dotted line at the spin-gap axis indicates it's value ($\sim 150 K$) corresponding to the equilibrium volume.  The spin-gap value is quite close to the experimental observation.\cite{kuo} The increase of spin-gap with volume-change is actually a manifestation of the variation of the couplings near equilibrium volume, which are shown at the inset of the upper panel of Figure 7.

Effect of size-confinement is also studied after confining a 36-atoms unit cell cluster and 72-atoms $1\times1\times2$ and $2\times1\times1$ supercell clusters and calculating the effective exchange integrals (Table III). A general observation not only indicates an increase of the exchange integrals but the geometric frustration of the bulk system seems to disappear as some such exchanges becomes ferromagnetic. So, in reduced dimension, the system may show a long-range ordering.

\begin{table}[tb]
\caption{\label{tab2} Relevant exchange couplings constants in meV for the Cu$_{2}$PO$_{4}$OH system under size-reduction.}
\begin{ruledtabular}
\begin{tabular}{ccccccc}
System & $J_{1}$ & $J_{1}^{\prime}$ & $J_{2}$ & $J_{2}^{\prime}$ & $J_{3}$ & $J_{4}$  \\
Cu$_{2}$PO$_{4}$OH & -35.56 & -35.56 & 20.32 & 20.32 & 16.407 & -10.69 \\
\end{tabular}
\end{ruledtabular}
\end{table}

\section{Conclusion}
In conclusion, for the general class of Libethenite system, the effect of substitutional impurity induced transformation from a three dimensional antiferromagnet to a low-dimensional quantum spin-1/2 tetramer system is investigated with the combined effects of structural and localized band-structure analysis along this series. The interplay between the structural distortion introduced by Cu-substitution at Co-site and its manifestation into the spin-magnetic structure of the series of systems are investigated with the help of appropriate DFT techniques. Identification of significant exchange paths and calculation of the corresponding exchange integrals provides the input to construct the model Hamiltonian for the low-dimensional system. All these studies have a close match with the earlier experimental observations. \section{Acknowledgement}
We would like to thank useful discussions with Alexander Yaresko and I. Bose. The authors would also like to acknowledge the Indo-EU project MONAMI and BARC-ANUPAM supercomputing facility.

\end{document}